\documentclass[superscriptaddress,reprint,10pt,a4paper,onecolumn,nolongbibliography]{revtex4-2}

\usepackage[bold-style=ISO]{unicode-math}
\makeatletter
\@ifpackageloaded{unicode-math}{\def\bm{\symbf}}{\usepackage{bm}}
\makeatother

\usepackage[fallback]{xeCJK}

\usepackage{fontawesome5}
\def\trash{\scalebox{.75}{\faTrash*}}

\usepackage{amsfonts}
\usepackage{siunitx}
\DeclareSIUnit{\angstrom}{\textup{\r A}}

\usepackage{graphicx}
\usepackage{xcolor}
\usepackage[normalem]{ulem}
\usepackage{makecell}
\usepackage{orcidlink}

\AtBeginDocument{
\let\Re\relax
\let\Im\relax
\DeclareMathOperator{\Re}{Re}
\DeclareMathOperator{\Im}{Im}
}
\renewcommand{\[}{\begin{equation}}
\renewcommand{\]}{\end{equation}}
\newcommand{\textsub}[2]{{#1}_{\text{#2}}}

\newcommand{\ii}{\mathrm{i}}
\newcommand{\dd}{\mathrm{d}}
\newcommand{\ee}{\mathrm{e}}
\newcommand{\EF}{E_{\text{F}}}
\DeclareMathOperator{\erf}{erf}
\DeclareMathOperator{\erfc}{erfc}

\DeclareMathOperator{\poly}{poly}
\DeclareMathOperator{\diag}{diag}
\DeclareMathOperator{\CQFT}{CQFT}

\def\RTE{\mathcal{R}}
\DeclareMathOperator{\Prep}{Prep}

\newcommand{\EQ}[1]{Eq.~(\ref{#1})}
\newcommand{\EQs}[1]{Eqs.~(\ref{#1})}
\newcommand{\FIG}[1]{Fig.~\ref{#1}}

\begin{document}

\title{Quantum algorithms for density functional theory with minimal readout}

\author{Yuansheng~Zhao\orcidlink{0000-0002-0362-4092}}
\email{yszhao@g.ecc.u-tokyo.ac.jp}
\affiliation{Quemix Inc., Chuo-ku, Tokyo 103-0027, Japan}
\affiliation{Department of Physics, The University of Tokyo, Bunkyo-ku, Tokyo 113-0033, Japan}
\author{Hirofumi~Nishi\orcidlink{0000-0001-5155-6605}}
\affiliation{Quemix Inc., Chuo-ku, Tokyo 103-0027, Japan}
\affiliation{Department of Physics, The University of Tokyo, Bunkyo-ku, Tokyo 113-0033, Japan}
\author{Taichi~Kosugi\orcidlink{0000-0003-3379-3361}}
\affiliation{Quemix Inc., Chuo-ku, Tokyo 103-0027, Japan}
\affiliation{Department of Physics, The University of Tokyo, Bunkyo-ku, Tokyo 113-0033, Japan}
\author{Satoshi~Hirose\orcidlink{0000-0002-6576-9721}}
\affiliation{Research Center of Excellence, Honda R\&D Co., Ltd., Shimotakanezawa 4630, Haga-machi, Haga-gun, Tochigi 321-3393, Japan}
\author{Hiroki~Sakagami}
\affiliation{Research Center of Excellence, Honda R\&D Co., Ltd., Shimotakanezawa 4630, Haga-machi, Haga-gun, Tochigi 321-3393, Japan}
\author{Tatsuki~Oikawa}
\affiliation{Research Center of Excellence, Honda R\&D Co., Ltd., Shimotakanezawa 4630, Haga-machi, Haga-gun, Tochigi 321-3393, Japan}
\author{Tatsuya~Okayama}
\affiliation{Research Center of Excellence, Honda R\&D Co., Ltd., Shimotakanezawa 4630, Haga-machi, Haga-gun, Tochigi 321-3393, Japan}
\author{Yu-ichiro~Matsushita\orcidlink{0000-0002-9254-5918}}
\affiliation{Quemix Inc., Chuo-ku, Tokyo 103-0027, Japan}
\affiliation{Department of Physics, The University of Tokyo, Bunkyo-ku, Tokyo 113-0033, Japan}
\affiliation{Quantum Materials and Applications Research Center, National Institutes for Quantum Science and Technology (QST), 2-12-1 Ookayama, Meguro-ku, Tokyo 152-8550, Japan}

\begin{abstract}
While quantum computers have shown significant promise for electronic structure calculations, their potential to accelerate density functional theory (DFT) calculations remains unclear.
In this work, we present a qubit-efficient encoding scheme for wavefunctions in Kohn--Sham (KS) DFT, together with a quantum algorithm that computes all occupied orbitals simultaneously. We further show that our algorithm is particularly well suited to the Harris functional, enabling the total energy to be evaluated with a potential exponential speedup over classical approaches by entirely avoiding the costly readout of the electronic density.
In addition, we propose a second method for achieving self-consistent DFT calculations using multiple copies of the wavefunction, which likewise circumvents density readout. The applicability of our algorithms is demonstrated through several numerical examples, and their efficiency is compared with that of existing approaches.
\end{abstract}

\maketitle

\section{Introduction}

Fast and accurate prediction of the electronic properties for a given material is one of the most crucial tasks in computational condensed matter physics and quantum chemistry.
However, the exact description of the many-electronic wavefunction costs exponential computational resource and quickly becomes intractable even for medium-sized molecules.
Therefore, to solve practical problems, mean-field methods such as Hartree--Fock \cite{hartree1928wave,fock1930fock,slater1930note} or Kohn--Sham (KS) density functional theory (DFT) \cite{hohenberg1964inhomogeneous,kohn1965self} are widely employed as standard approaches in first-principles calculations.
Here, the electronic wavefunctions are restricted to those expressible by a single Slater determinant, and the computational cost typically scales only with $\mathcal O (\textsub N{atom}^3)$ where $\textsub N{atom}$ is the number of atoms, enabling calculations with $10^3$ atoms on supercomputers.
In the meantime, the development of quantum computers has demonstrated great potential to solve the true many-electronic wavefunctions that are impossible for classical computers \cite{abrams1997simulation,babbush2023quantum}, and in fact quantum chemistry is one of the most promising applications of the quantum computers.
It is therefore of interest to use quantum computers to accelerate the DFT calculation.

While quantum computers have been demonstrated to be useful in several aspects of DFT calculation such as generating KS potentials \cite{baker2020density},
the possibility of directly solving the KS equation on quantum computers is still unclear and faces several challenges.
Firstly, an efficient encoding of KS orbitals needs to be devised.
Most of the quantum algorithms so far encode the genuine many-body wavefunctions $\psi(\bm r_1,\cdots,\bm r_N)$ \cite{kosugi2023exhaustive}.
This corresponds to the exact wavefunction of the interacting electrons, but is not well compatible with the single Slater determinant used in KS-DFT.

Secondly, quantum solvers for the Hermitian eigenvalue problems, such as probabilistic imaginary time evolution (PITE) method \cite{kosugi2022imaginary}, generally only work for the ground state and cannot be generalized to higher states in a simple manner.
However, in DFT calculation of a $\textsub N{elec}$-electron system, all the lowest $\gtrsim\frac12\textsub N{elec}$ eigenstates of the KS Hamiltonian must be solved simultaneously, and an efficient solver for such multiple-state case is required.
While this difficulty can be circumvented by using orbital-free DFT (OF-DFT) \cite{witt2018orbital} which is another scheme of DFT different from KS-DFT, 
allowing usual quantum solvers can be directly employed \cite{nishiya2024orbital}, the accuracy OF-DFT itself is usually not sufficient for production level.
For KS-DFT, quantum algorithms based on eigenvalue filtering \cite{ko2023implementation} or ensemble VQE \cite{senjean2023toward} have also been proposed, but their computational efficiency is still limited (see section \ref{sec:discussion} below).

Finally, as the KS Hamiltonian depends on the unknown electronic density, the DFT computation requires a self-consistency field (SCF) solution.
On quantum computers, this would typically require repeated reading out of the calculated electron density from quantum registers and re-encoding of the KS Hamiltonian.
Because the quantum state collapses upon measurement, this readout of electronic density would require $\mathcal O(\textsub N{elec})$ shots, which heavily limits the performance of DFT on quantum computers.
However, none of the existing algorithms have satisfactorily solved this problem.

In this article, we develop several quantum algorithms for KS-DFT calculation which minimize the number of density readouts.
It is demonstrated that Harris functional \cite{harris1985simplified,foulkes1989tight}, a non-SCF version of KS functional with very high accuracy, is highly suitable for quantum computing where the density readout can be completely eliminated, and its non-SCF error may be minimized using a variational method.
We then propose a qubit-efficient encoding that is well suitable for the description of wavefunctions in DFT, and a quantum algorithm to calculate all the KS orbitals at once.
Additionally, we show that SCF calculation can also be achieved without any density readout by using copies of wavefunctions.
The efficiency of our quantum algorithms is demonstrated using several numerical experiments and we finally discuss the computational efficiency of our algorithms. 

\section{Quantum algorithms for DFT}
\subsection{Kohn--Sham density functional theory}
In KS-DFT \cite{hohenberg1964inhomogeneous,kohn1965self}, the energy of the electronic system is expressed by a functional of electronic density $\rho(\bm r)$ 
\[\label{eq:ksfunct}\textsub E{KS}[\rho]=T[\rho]+\int \rho(\bm r)\textsub V{ext}(\bm r) \,\dd^3\bm r + \textsub EH[\rho] +\textsub E{XC}[\rho].\]
Here, $T[\rho]$ and $\textsub V{ext}(\bm r)$ denote the kinetic energy and external potential (exerted by nucleus), respectively; $\textsub EH[\rho]=\frac12\iint \frac{\rho(\bm r)\rho(\bm r')}{|\bm r-\bm r'|}\,\dd^3\bm r\,\dd^3\bm r'$ is the Hartree energy and finally $\textsub E{XC}[\rho]$ is the exchange-correlation (XC) energy.
The ground-state electronic density is determined by minimizing this energy functional under the constraint of constant number of electrons $\textsub N{elec}$, i.e., $\int \rho(\bm r) \,\dd^3\bm r=\textsub N{elec}$, and is equivalent to self-consistently solving the KS equation, given by
\[\label{eq:ks}\textsub H{KS}[\rho]\psi_i^{\text{KS}}(\bm r) \equiv \left[\underbrace{-\frac12\nabla^2}_T +\underbrace{\textsub V{ext}(\bm r) +\textsub V{H}[\rho](\bm r) +\textsub V{XC}[\rho](\bm r)}_{\textsub V{KS}[\rho](\bm r)} \right]\psi_i^\text{KS}(\bm r) = \varepsilon_i \psi_i^{\text{KS}}(\bm r),\]
with
\[\label{eq:rhodef}\rho(\bm r)=2\sum_{\text{occupied}}|\psi_i^\text{KS}(\bm r)|^2.\]
Here, $\psi_i^\text{KS}$ is the $i$-th single-electronic KS orbital with energy $\varepsilon_i$, $\textsub V{H}[\rho](\bm r)=\int \frac{\rho(\bm r)\rho(\bm r')}{|\bm r-\bm r'|}\,\dd^3\bm r'$ and $\textsub V{XC}[\rho](\bm r)=\frac{\delta \textsub E{XC}[\rho]}{\delta \rho(\bm r)}$ are the Hartree and XC potential, respectively, and the factor 2 in \EQ{eq:rhodef} comes from spin degeneration.
Once the KS equation has been solved, the total energy can then be calculated by \EQ{eq:ksfunct}, or equivalently by
\[\label{eq:ksetot}\textsub E{KS}=2\sum_{\text{occupied}} \varepsilon_i - \textsub EH[\rho]- \int \textsub V{XC}(\bm r)\rho(\bm r)\,\dd^3\bm r +\textsub E{XC}[\rho].\]
Here, the first and remaining terms are referred to as ``band energy'' and ``double counting'', respectively.

\subsection{Avoiding density readout by Harris functional}

\subsubsection{Harris functional}

As discussed above, the density readout is an expensive operation in quantum computing, but this is required for SCF calculation because the  $\textsub H{KS}$ depends on the electronic density which is unknown \textit{a priori}.
Moreover, even if the final KS orbitals have been successfully prepared on the quantum registers, 
to compute the total energy using \EQ{eq:ksetot}, the double counting terms (the second to fourth terms) would still require the density to be readout.
Such problem is one of the largest obstacles to performing DFT on quantum computers.
However, we have found that this can be all overcome by employing the Harris functional \cite{harris1985simplified,foulkes1989tight},
which is initially proposed as a non-SCF DFT scheme intended for high-speed calculations.
Contrary to the KS scheme which requires a SCF loop, the KS Hamiltonian here is only solved once at some input electronic density $\textsub \rho{in}$,
often the sum of density of isolated atoms.
The total energy is then given by
\[\label{eq:ergharris}\textsub E{Harris}[\textsub \rho{in}]=2\sum_{\text{occupied}} \varepsilon_i - \textsub EH[\textsub \rho{in}]-\int \textsub V{XC}[\textsub \rho{in}](\bm r) \textsub \rho{in}(\bm r)\,\dd^3\bm r +\textsub E{XC}[\textsub \rho{in}].\]
This is at first glance the same as \EQ{eq:ksetot}, but the difference is that the double-counting terms in $\textsub E{Harris}$ only involves the input density, instead of the density from the one-shot solution of KS equation.
It can be shown that the error in the Harris functional is only second order in the density difference $\mathcal O[(\textsub \rho{in}-\textsub\rho{KS})^2]$ where $\textsub\rho{KS}$ denotes the SCF electronic density, and in practice it also demonstrates high accuracy even for covalent systems, being able to reproduce the energy difference between cubic and hexagonal silicon from our tests.

In classical computing, Harris functional unfortunately only provides limited speedup because the band energy still requires a diagonalization of the Hamiltonian anyway, which scales with $\mathcal O(\textsub N{atom}^3)$. 
However, in the context of quantum computing, the band energy can be computed efficiently as discussed below, and will no longer become a bottleneck.
More importantly, as the Harris functional does not require an SCF loop (the Hamiltonian is already known) and the double-counting terms in the total energy \EQ{eq:ergharris} only involve the input electron density (also known in advance, and can be computed on classical computers), the problematic density readout in quantum computing is totally unnecessary and can be completely avoided.
In other words, the simplification brought by Harris functional perfectly fits the need for quantum computers, and a significant acceleration compared to classical DFT calculations can be achieved.

\subsubsection{$\bm k$-point sampling}

For periodic system, it is also possible to perform $\bm k$-point sampling under Harris functional in quantum computing.
Since the wavefunctions at different $\bm k$ points belong to independent eigen systems,
the calculation at each $\bm k$ point is independent from each other and can be performed sequentially, one $\bm k$ point at a time.
Here, the KS potential $\textsub V{KS}$ in \EQ{eq:ks} is the same across different $\bm k$ points, and the only difference is that the kinetic energy, which in momentum space reads $\diag \left(\frac{|\bm G+\bm k|^2}2\right)$ with $\bm G$ being reciprocal vectors of the simulation cell.
Therefore, the quantum circuits for computation at one $\bm k$ point can be easily adapted for calculation at other $\bm k$ points.

\subsubsection{Variational method for improving accuracy}\label{sec:harrisvar}

When the true electron density is very different from the density sum of isolated atoms (used as $\textsub \rho {in}$),
the Harris functional may not be accurate enough and improving its accuracy in this case is highly desirable.
Here we propose a simple yet effective variational method for such cases.
While the KS functional [\EQ{eq:ksfunct}] is variational (since the true energy is obtained by minimizing this functional),
the Harris functional has been proven to be a saddle point or local minimum at the $\textsub \rho{in}=\textsub \rho{KS}$ \cite{robertson1991does} and thus does not permit a strict variational approach.
However, it is in practice ``almost always'' found to be anti-variational \cite{read1989tests,finnis1990harris}, i.e., the energy will never be higher than the true energy, unless very unphysical $\textsub\rho{in}$ are used in the calculation.
Therefore, a practically useful method to improve the accuracy is to adjust $\textsub\rho{in}$ and maximize the total energy.
This method is similar to the variational quantum eigen solver (VQE) and can be useful on quantum computer given that the total energy of Harris functional can be efficiently computed.
For ionic compound, an obvious yet effective parameter to adjust is the charge of each atom, and high-accuracy total energies can be computed in this way from our numerical tests in subsection \ref{sec:lihHarris}.

\subsection{Qubit-efficient encoding of wavefunctions}

Here, we describe our encoding scheme of wavefunctions for DFT calculation on quantum computers.
As in conventional first-quantized formalism, 
to solve the KS equation, we discretize the simulation cell of size $L_x\times L_y \times L_z$ into $\textsub N{grid}=2^{n_x}\times 2^{n_y} \times 2^{n_z}$ uniform grid, and the grid point $\bm r^{ijk}=(i L_x/2^{n_x},j L_y/2^{n_y},k L_z/2^{n_z})$. Using $\textsub n{grid}=n_x+n_y+n_z$ qubits and computational basis $|\bm r\rangle_{\textsub n{grid}}\equiv |i\rangle_{n_x}|j\rangle_{n_y}|k\rangle_{n_z}$, a one-electron wavefunction can be encoded as \[|\psi\rangle=\sqrt{\Delta V}\sum_{\bm r} \psi(\bm r) |\bm r\rangle_{\textsub n{grid}}.\]
Here, $\Delta V=\frac{L_x L_y L_z}{\textsub N{grid}}$ and the wavefunction is normalized that $\int |\psi(\bm r)|^2\,\dd^3 \bm r \approx \sum_{\bm r} |\psi(\bm r)|^2\Delta V =1$.

To encode $\textsub N{band}$ orbitals simultaneously, we employ a qubit-efficient (QE) encoding scheme.
In addition to the $\textsub n{grid}$ grid qubits, we introduce $\textsub n{band}=\lceil \log_2 \textsub N{band} \rceil$ label qubits, and encode the whole wavefunction for DFT calculation as
\[|\psi_N\rangle=\sqrt{\frac{\Delta V}{\textsub N{band}}}\sum_{i\bm r}\psi_i^\text{KS}(\bm r) |i\rangle_{\textsub n{band}}|\bm r\rangle_{\textsub n{grid}}.\] 
This encoding is termed ``qubit-efficient'' because it only uses $\textsub n{band}+\textsub n{grid}$ qubits, which is the minimum number to store all the $\textsub N{band}\textsub N{grid}$ complex numbers, and exponentially reduced from the usual $\textsub N{band}\textsub n{grid}$ that is required for conventional encoding of many-body wavefunction.

In order to accommodate $\textsub N{elec}$ electrons, the minimum value for $\textsub N{band}$ is evidently $\frac12\textsub N{elec}$.
However, as with DFT calculation on classical computers, using a larger value for $\textsub N{band}$ (i.e., including some unoccupied bands) is also possible,
and actually may be beneficial for computational accuracy and stability, as discussed in subsection \ref{sec:cross} below. 
Additionally, one should note that this encoding does not automatically ensure that the encoded KS orbitals are orthonormal to each other, and this constraint needs to be carefully imposed during the initialization and computation.

\begin{figure}
    \includegraphics[scale=0.55]{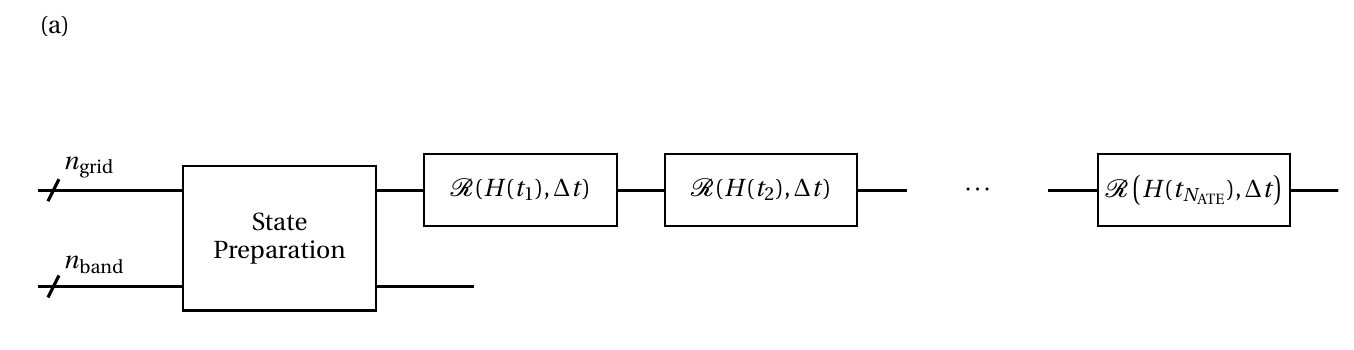}\\
    \includegraphics[scale=0.55]{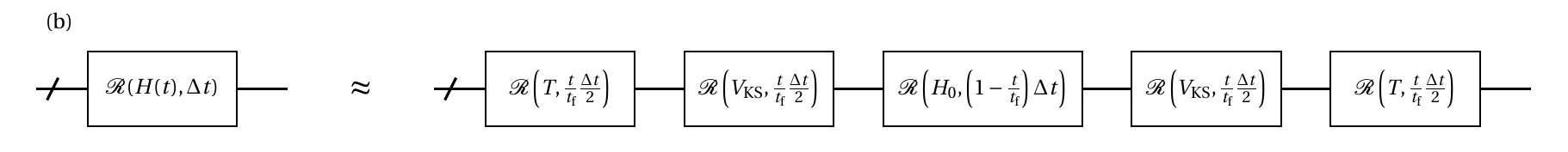}\\
    \includegraphics[scale=0.55]{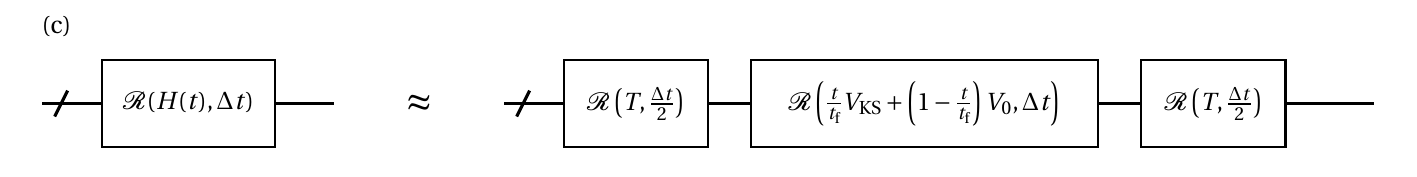}\\
    \caption{
    (a) Quantum circuits for performing adiabatic real-time evolution (ATE) with qubit-efficient encoding. Here $\RTE(H,\Delta t)=\ee^{-\ii\Delta t H}$ and $\textsub N{ATE}$ denotes the real time evolution operator and total number of steps, respectively.
    (b) The implementation of $\RTE(H(t),\Delta t)$ with Suzuki--Trotter splitting in ATE for a general $H_0$. Here $T$ and $\textsub V{KS}$ denote the kinetic energy and Kohn--Sham potential operator, respectively, and $\textsub tf$ is the total time duration for ATE.
    (c) The implementation of $\RTE(H(t),\Delta t)$ when $H_0$ can be written as $T+V_0$.
    }\label{fg:circ}
\end{figure}

\subsection{Quantum solvers for multiple eigen states}
\subsubsection{Adiabatic real-time evolution}

As the lowest $\textsub N{band}$ eigen states of the KS Hamiltonian must all be solved, 
the PITE algorithm that can efficiently solve the lowest eigen-state is in general not applicable here because all orbitals will collapse to the lowest state.
Instead, we have found that the adiabatic real-time evolution (adiabatic RTE, ATE) algorithm \cite{nishiya2024first} can be employed to compute all the higher states and works with our QE encoding. 
Specifically, we consider a trivially-solvable Hamiltonian $H_0$ and initialize the wavefunction to be its lowest $\textsub N{band}$ eigenstates $\psi_i^0$, i.e. $|\psi_N(t=0)\rangle=\sqrt{\frac{\Delta V}{\textsub N{band}}}\sum_{i\bm r}\psi_i^0(\bm r) |i\rangle|\bm r\rangle$.
The state on grid qubits is then evolved in real time from $t=0$ to final time $\textsub tf$, under a time-dependent Hamiltonian $H(t)$ which only acts on the grid qubits and satisfies $H(0)=H_0$ and $H(\textsub tf)=\textsub H{KS}$, as shown in \FIG{fg:circ} (a).
From the adiabatic theorem \cite{sakuraiQM}, if the eigen-energies of $H(t)$ are not degenerate at any time and $\partial H(t) /\partial t$ is sufficiently small, the final state will be just the lowest $\textsub N{band}$ eigen-state of $\textsub H{KS}$, i.e. $|\psi_N(t=\textsub tf)\rangle=\sqrt{\frac{\Delta V}{\textsub N{band}}}\sum_{i\bm r}\psi_i^\text{KS}(\bm r) |i\rangle|\bm r\rangle$.
Importantly, because the time-evolution operator is always unitary even if the adiabaticity condition is violated, the overlap matrix $S_{ij}\equiv\langle\psi_i|\psi_j\rangle$ is a constant independent of time. This indicates that if the initial wavefunctions are mutually orthonormal, so will be the final states, and no extra orthogonalization step will be necessary.

Here, we here use linear interpolation \[\label{eq:htate}H(t)=\left(1-\frac t{\textsub tf}\right)H_0+ \frac t{\textsub tf}\textsub H{KS},\]
and discretize the time interval into $\textsub N{ATE}$ steps, i.e.,
\[|\psi_N(\textsub tf)\rangle= \left(\prod_{i=\textsub N{ATE}}^1 \RTE(H(t_i),\Delta t)\right)|\psi_N^0\rangle,\] 
where $\RTE(H,\tau)=\ee^{-\ii\tau H}$ is the time-evolution operator, $\Delta t=\textsub tf/\textsub N{ATE}$ and $t_i=\left(i-\frac12\right)\Delta t$.
The $\RTE(H(t_i),\Delta t)$ operator can be implemented by several methods \cite{childs2018toward} and an efficient implementation is crucial for the performance of ATE.
We here employ the second order Suzuki--Trotter (ST) splitting \cite{suzuki1976generalized} scheme as shown in \FIG{fg:circ} (b) for a general $H_0$,
\[\label{eq:stdecomp}\begin{split}
\RTE(H(t_i,\Delta t)\approx {}&{}\exp\left(-\ii \frac {t_i}{\textsub tf}\frac{\Delta t}2T\right) \cdot \exp\left(-\ii \frac {t_i}{\textsub tf}\frac{\Delta t}2\textsub V{KS}\right) \cdot \exp\left(-\ii \left(1-\frac {t_i}{\textsub tf}\right)\Delta t H_0\right)\cdot{}\\ &{}\exp\left(-\ii \frac {t_i}{\textsub tf}\frac{\Delta t}2\textsub V{KS}\right) \cdot\exp\left(-\ii \frac {t_i}{\textsub tf}\frac{\Delta t}2T\right),
\end{split}\]
where $T$ and $\textsub V{KS}$ denote the kinetic energy and KS potential operators, respectively.
If the $H_0$ can also be written as the sum of kinetic and potential terms, i.e., $H_0=T+V_0$, we can use a different type of ST decomposition from \EQ{eq:stdecomp},
\[\label{eq:stdecomp2}
\RTE(H(t_i,\Delta t)\approx\exp\left(-\ii \frac{\Delta t}2T\right) \cdot \exp\left[-\ii \Delta t\left(\left(1-\frac {t_i}{\textsub tf}\right) V_0+\frac {t_i}{\textsub tf} \textsub V{KS}\right)\right]\cdot\exp\left(-\ii \frac{\Delta t}2T\right),\]
as shown in \FIG{fg:circ} (c).

The kinetic operator is diagonal $\frac{|\bm G+\bm k|^2}2$ is momentum space and therefore $\exp(-\ii \tau T)$ is implementable by using the centered quantum Fourier transform (CQFT) \cite{nishiya2024first} as 
\[\exp(-\ii \tau T)= \CQFT \cdot \exp \left[-\ii\tau \diag \left(\frac{|\bm G+\bm k|^2}2\right)\right]\cdot \CQFT^\dagger.\]
Both CQFT and quadratic function can be implemented by circuits with depth $\mathcal O(\textsub n{grid}^2)$ \cite{benenti2008quantum}. 
$\textsub V{KS}$ is diagonal in real space for local potentials, but due to its complex behavior as a function of $\bm r$, it is rather difficult to implement.
While the potential from a single nucleus can be implemented with only $\mathcal O[\poly(\textsub n{grid})]$ in circuits depth \cite{huang2025approximate}, 
implementing the potential from all $\textsub N{atom}$ nuclei cannot be easily parallelized and will cost $\mathcal O (\textsub N{atom})$ in circuit depth in general.
However, with additional $\mathcal O (\textsub N{grid})$ ancilla bits, the circuit depth can be reduced to $\mathcal O(\textsub n{grid})$ by employing the method described in Ref.~\cite{nishiya2024orbital}, thus possibly enabling a exponential speed up compared to the classical algorithms.

A good initial Hamiltonian $H_0$ is also very important for ATE.
Since the adiabaticity requires $\partial H(t) /\partial t$ to be small, an $H_0$ close to $\textsub H{KS}$ is very beneficial to reduce $\textsub tf$ and the total circuit depth.
In the meantime, efficient implementation of $\RTE(H_0,\tau)$ and preparation of its eigenstates are required.
Here, considering that some state preparation circuit $\Prep: |i\rangle_{\textsub n{grid}}\mapsto \sqrt{\Delta V}\sum_{\bm r} \psi_i^0(\bm r)|\bm r\rangle_{\textsub n{grid}}$ $(i=0,\cdots,\textsub N{band}-1)$ 
is required to initialize the wavefunctions anyway,
we can choose \[\label{eq:h0}H_0= -E_0\sum_i | \psi_i^0 \rangle\langle \psi_i^0|,\]
and the RTE operator can then be implemented by using generalized phase operator as 
\[\RTE(H_0,\tau)=\Prep \cdot \exp \left[-\ii\tau E_0\diag (\underbrace{1,\cdots ,1}_{\textsub N{band}},0,\cdots ,0)\right]\cdot \Prep^\dagger.\]

\subsubsection{Result readout}

\begin{figure}
    \includegraphics[scale=0.55]{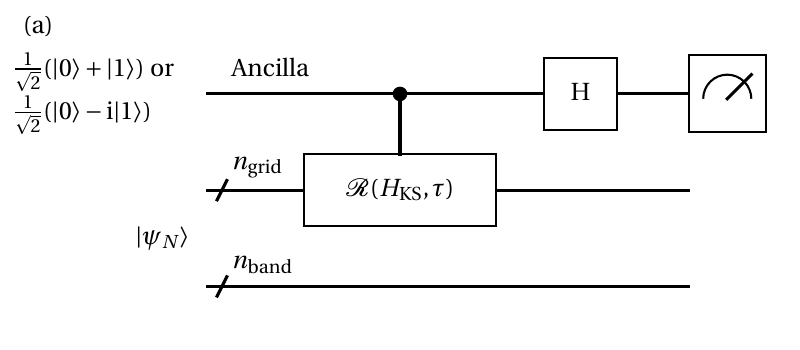}
    \includegraphics[scale=0.55]{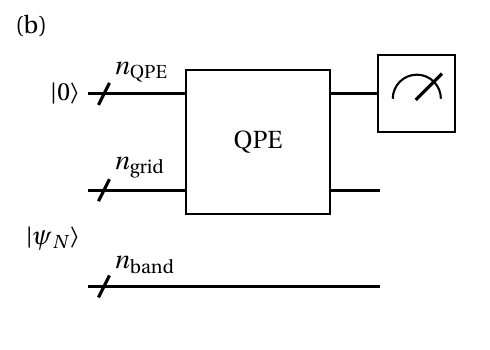}
    \caption{(a) Quantum circuit for reading out the band energy using Hadamard test. The real and imaginary part of $\sum_i \langle\psi_i^\text{KS}|\RTE(\textsub H{KS},\tau)|\psi_i^\text{KS}\rangle$ can be obtained when the ancilla bit is initialized to be $\frac1{\sqrt2}(|0\rangle+|1\rangle)$ and $\frac1{\sqrt2}(|0\rangle-\mathrm i |1\rangle)$, respectively. (b) The circuit for reading out the density of states via quantum phase estimation (QPE).}\label{fg:read}
\end{figure}

After the final state $|\psi_N(\textsub tf)\rangle=\sum_{i\bm r}\psi_i^{\text{KS}}(\bm r)|i\rangle|\bm r\rangle$ has been prepared on quantum registers, 
we can then evaluate the total energy using \EQ{eq:ksetot}.
We here only focus on the band energy $2\sum_{\text{occupied}} \varepsilon_i$ since the electronic density readout is not required in Harris functional and can be efficiently evaluated classically (the electronic density can be nonetheless accessed by reading the grid qubits).

When there is no unoccupied bands in the calculation (i.e., $\textsub N{band}=\frac12\textsub N{elec}$), the easiest way is to use the Hadamard test of $\RTE(\textsub H{KS},\tau)$ in \FIG{fg:read} (a), where $\tau$ is chosen to be small enough such that $\tau\ll 1/(\max_{\text{occupied}} \varepsilon_i-\min \varepsilon_i)$.
When the ancilla bit is initialized to be $\frac1{\sqrt2}(|0\rangle+|1\rangle)$ or $\frac1{\sqrt2}(|0\rangle-\mathrm i |1\rangle)$,
the probability of obtaining 0 in the measurement is $\frac12+\frac1{2\textsub N{band}} \sum_i \Re \langle\psi_i^\text{KS}|\RTE(\textsub H{KS},\tau)|\psi_i^\text{KS}\rangle$ or $\frac12+\frac1{2\textsub N{band}} \sum_i \Im \langle\psi_i^\text{KS}|\RTE(\textsub H{KS},\tau)|\psi_i^\text{KS}\rangle$, respectively.
We can then compute the band energy by 
\[2\sum_{i} \varepsilon_i\approx \frac 2{\tau} \Im \log \left(\sum_i \langle\psi_i^\text{KS}|\RTE(\textsub H{KS},\tau)|\psi_i^\text{KS}\rangle\right).\]
Importantly, the number of shots to reach a prescribed accuracy $\delta$ in energy \emph{per atom} is $\mathcal O\left(\frac{\max_{\text{occupied}} \varepsilon_i-\min \varepsilon_i}{\delta^2}\right)$ and does not directly depend on the system size, making the algorithm highly favorable for large systems. The $\delta^{-2}$ dependence may be reduced to $\delta^{-1}$ by quantum amplitude amplification \cite{brassard2000quantum}.

In case when the density of states (DOS) is required alongside the energy, or when some unoccupied bands are included in the calculation, we instead employ the quantum phase estimation (QPE) \cite{kitaev1995quantum} of the $\RTE(\textsub H{KS},\tau)$ operator.
As shown in \FIG{fg:read} (b), with $\textsub n{QPE}=\log_2 \textsub N{QPE}$ working qubits, the results of QPE before measurement is
\[\label{eq:qpestate}
|\text{QPE}\rangle=\sum_{ki\bm r}\sqrt{\frac{\Delta V}{\textsub N{band}}}\psi_i^{\text{KS}}(\bm r) F_{ik} |k\rangle|i\rangle|\bm r\rangle.
\]
Here, $|k\rangle$ denotes the basis of QPE working qubits, and $F_{ik}=\frac{\exp\left[\ii \textsub N{QPE}\left(\frac{2\pi k}{\textsub N{QPE}}-\Delta t\varepsilon_i\right)\right]-1}{\textsub N{QPE}\left[\exp\left[\ii\left(\frac{2\pi k}{\textsub N{QPE}}-\Delta t\varepsilon_i\right)\right]-1\right]}$, which is sharp-peaked near $\frac{2\pi k}{\textsub N{QPE}\Delta t}=\varepsilon_i$ for large $\textsub N{QPE}$.
The probability of obtaining $k$ in the measurement is then
\[\label{eq:qpeprob}
\Pr(k)=\frac{1}{\textsub N{band}}\sum_i |F_{ik}|^2\approx \frac{1}{\textsub N{band}} D\left(\frac{2\pi k}{\textsub N{QPE}\Delta t}\right).\] 
where $D$ denotes the DOS of the lowest $\textsub N{band}$ states.
With the obtained DOS, we can determine the Fermi level $\EF$ by solving
\[\label{eq:fermi}\int 2D(\varepsilon)f(\varepsilon-\EF)\,\dd\varepsilon=\textsub N{elec}\]
where $f$ is the smearing function. 
The band energy is then calculated as
\[\label{eq:fermie}2\sum_{\text{occupied}} \varepsilon_i =\int 2 \varepsilon D(\varepsilon)f(\varepsilon-\EF)\,\dd\varepsilon.\]
As the DOS is approximated by the histogram of the measurement results, the integrations in \EQs{eq:fermi} and (\ref{eq:fermie}) are effectively performed using Monte Carlo method.
Therefore, the number of shots to reach a designated accuracy $\delta$ on per-atom energy is $\mathcal O \left( \frac{\textsub N{band}}{\textsub N{elec}/2}\frac{\sigma (\text{DOS})}{\delta^2}\right)$ where $\sigma (\text{DOS})$ denotes the standard deviation of DOS and the first term $\frac{\textsub N{band}}{\textsub N{elec}/2}$ is due to the fact that only occupied states contribute to the total energy.
Similar to the method of Hadamard test above, the number of shots is also independent of the system size to reach a fixed accuracy of per-atom energy.

\subsubsection{Level crossing during ATE}\label{sec:cross}

The ATE requires the adiabatic condition to be satisfied during time evolution.
However, this is not always easy to ensure in practice, especially for metallic systems.
When level crossing (or near crossing) occurs, the orbitals in question will get hybridized and the eigenstates of $\textsub H{KS}$ will no longer be obtainable.
However, if hybridization occurs within the occupied subspace, it will have no effect on any calculated physical properties as in classical DFT calculations.
We have found that even the problematic case of crossing between occupied and unoccupied orbitals can be solved by increasing $\textsub N{band}$ in the calculation.
It can be shown (see Appendix \ref{seca:cross}) that the correct total energy can be obtained by QPE [\FIG{fg:read} (b)] if \[\label{eq:uijsqr}\sum_{i>\textsub N{band}}|\mathcal U_{ij}|^2\ll 1\] is satisfied for all occupied orbitals $j$, where $\mathcal U_{ij}=\langle\psi_j^{\text{KS}}|\mathcal U|\psi_i^0\rangle$ and $\mathcal U$ is the time-evolution operator of $H(t)$ from $t=0$ to $\textsub tf$.
Since the non-adiabatic coupling between level $i$ and $j$ scales with $\frac{\langle\psi_i(t)|\dot H|\psi_j(t)\rangle}{\varepsilon_i-\varepsilon_j}$ \cite{sakuraiQM,kato1950adiabatic,sarandy2004consistency}, 
the amplitude leakage is only significant between states whose energies are close. 
When $\textsub N{band}$ is taken to be sufficiently large so that $\varepsilon_i\gg \varepsilon_j$ for any $i>\textsub N{band}$ and occupied orbital $j$, 
one can expect $\mathcal U_{ij}$ to be very small between these states, i.e., \EQ{eq:uijsqr} will be satisfied.
In the limiting case of $\textsub N{band}=2^{\textsub n{grid}}$, i.e., including all the eigenstates of the Hamiltonian in the calculation, \EQ{eq:uijsqr} will be evidently satisfied automatically.
This also indicates that a practical way to validate the calculation results is to increase the $\textsub N{band}$ and check whether the result has converged. 
However, as the number of measurement shots required is proportional to $\textsub N{band}$ for fixed system size, $\textsub N{band}$ should be kept as small as possible in practice.

\subsection{Achieving self-consistency using copies}\label{sec:trace}

\begin{figure}
    \includegraphics[scale=0.55]{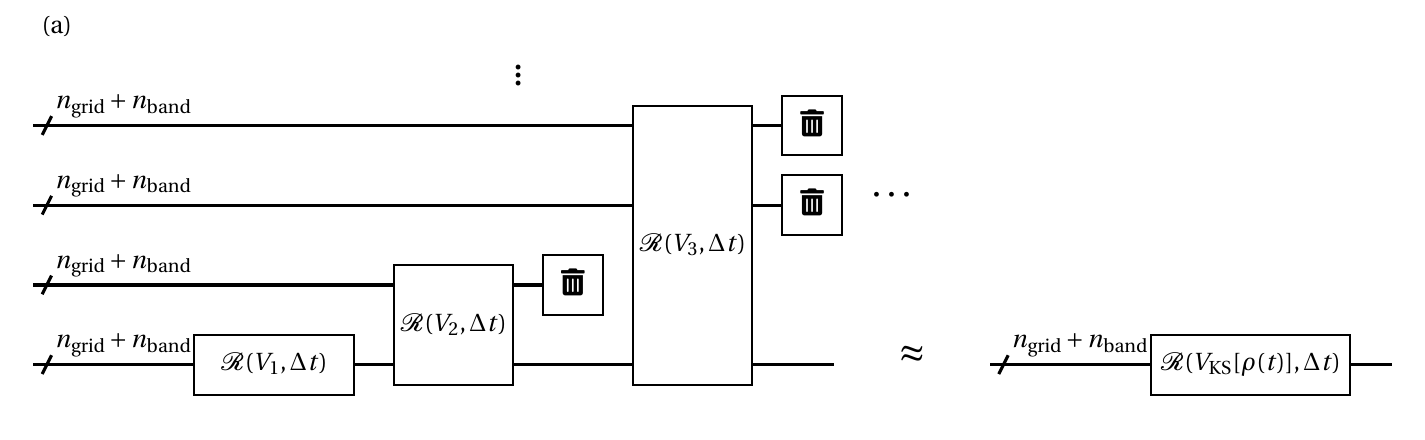}\\
    \includegraphics[scale=0.55]{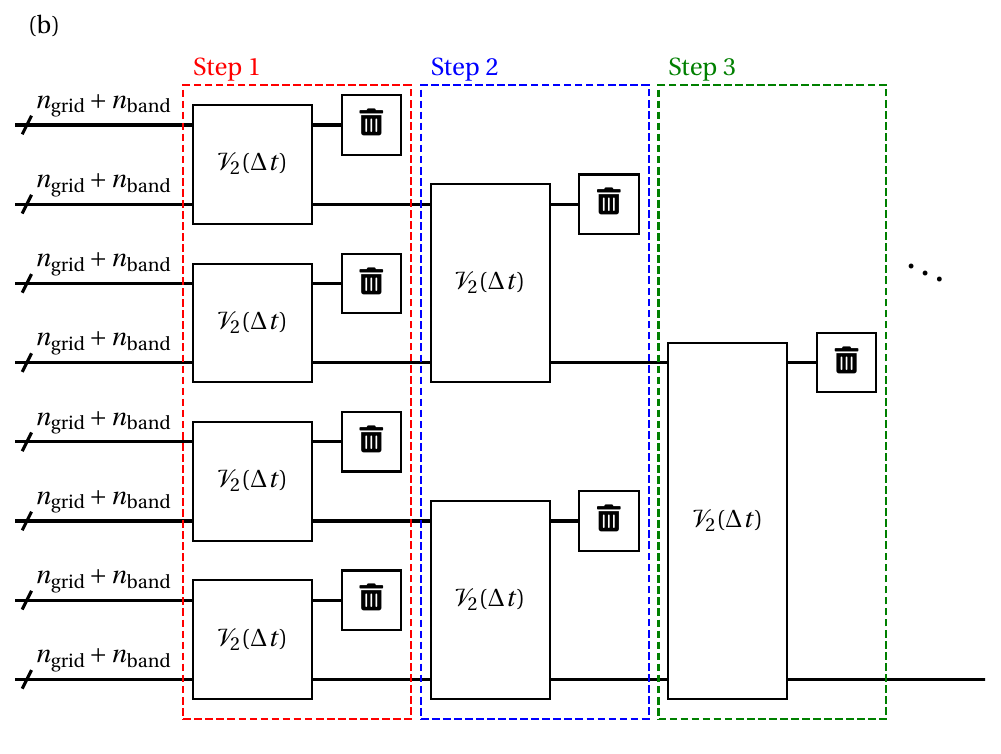}\\
    \includegraphics[scale=0.55]{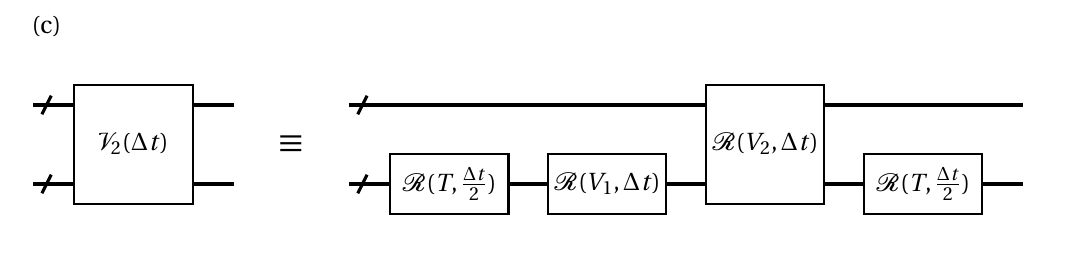}
    \caption{(a) The quantum circuit for performing real-time evolution of $\textsub V{KS}[\rho]$ of unknown $\rho$, but without reading it out, where the trash bin symbol ``\trash'' denotes tracing out and the operator $V_n$ is given by \EQ{eq:nbody} of the main text.
    The approximation becomes more accurate as higher order terms (omitted here by ``$\cdots$'') are included.
    (b) The circuit for performing self-consistency field calculation within ATE, where only up to the second order terms $V_2$ are included, and the circuit of the sub-routine $\mathcal V_2$ is given in (c).
    }\label{fg:circscf}
\end{figure}

Apart from pairing near perfectly with the Harris functional, our QE encoding can also be employed to perform the SCF calculations without density readout if the local density approximation (LDA) \cite{dirac1930note,perdew1992accurate} is employed for the XC energy and $\textsub N{band}=\frac12\textsub N{elec}$ (no unoccupied bands).
Within the framework of ATE, SCF can be achieved by making $\textsub H{KS}$ in \EQ{eq:htate} dependent on time as $\textsub H{KS}[\rho(t)]$ with $\rho(t)=2\sum_i |\psi_i(t)|^2$, i.e., evaluated on the (unknown) wavefunctions during ATE.

The $\rho$-dependence of $\textsub V{KS}$ reflects the Coulomb repulsions between electrons and makes the time evolution non-linear.
To incorporate such effects using the QE encoding, we expand the $\textsub V{KS}$ on the grid points using the series $\textsub V{KS}^{(n)}$ as 
\[\label{eq:mfn}\textsub V{KS}[\rho](\bm r_1)
\approx\sum_n  \textsub V{KS}^{(n)}[\rho](\bm r_1)
\equiv \sum_n \sum_{\bm r_2,\cdots,\bm r_\alpha} v_n(\bm r_1,\bm r_2,\cdots,\bm r_\alpha)\prod_{\alpha=2}^n\rho(\bm r_\alpha)\Delta V^{n-1}.
\]
with
\begin{align}
v_1(\bm r_1)&{}=\textsub V{ext}(\bm r_1) \label{eq:v1} \\
v_2(\bm r_1,\bm r_2)&{}=\frac1{|\bm r_1-\bm r_2|}+\frac{2c_2\delta(\bm r_1,\bm r_2)}{\Delta V}\label{eq:v2} \\
v_\alpha(\bm r_1,\cdots,\bm r_\alpha)&{}=\frac{\alpha c_\alpha\delta(\bm r_1,\cdots,\bm r_\alpha)}{\Delta V^{\alpha-1}}\quad\text{for }\alpha\ge 3,
\end{align}
where $\delta$ is the Kronecker delta (equals to 1 if all arguments are equal and 0 otherwise) and 
$c_\alpha$ is obtained by the polynomial approximation of the LDA XC energy $\rho \textsub \varepsilon{XC}(\rho)\approx \sum_\alpha c_\alpha \rho^\alpha$.

Non-linear operations in quantum computing can be achieved by using multiple copies of the wavefunctions \cite{lloyd2020quantum,leyton2008quantum,holmes2023nonlinear}.
In our case, we introduce $|\psi_N\rangle^n\equiv\bigotimes\nolimits^n|\psi_N\rangle$,
and the $n$-body interaction operator
\[\label{eq:nbody}V_n=\sum_{\{i_\alpha\bm {r}_\alpha\}} \textsub N{elec}^{n-1}v_n(\{\bm r_n\}) \bigotimes_{\alpha=1}^n |\bm r_\alpha \rangle|i_\alpha\rangle\langle i_\alpha|\langle\bm r_\alpha|\]
where $i_\alpha\bm r_\alpha$ is the index of the $\alpha$-th copy.
One can verify that 
\[\sum_{\bm r_1} \textsub V{KS}[\rho](\bm r_1)\rho(\bm r_1)\Delta V=\frac1{\textsub N{elec}}\sum_n {}^n\langle \psi_N|V_n|\psi_N\rangle^n,\]
i.e., the correct expectation values of $\textsub V{KS}$ can be represented using the $V_n$ operators.

However, the many-body operator above are different from the true $\textsub V{KS}[\rho(t)]$ operator which is inherently a one-body operator, and the required $\RTE(\textsub V{KS}[\rho(t)],\Delta t)$ in ATE is also not directly achievable by $\RTE(V_n,\Delta t)$.
Nonetheless, we note that the reduced density matrix of the first copy in the $n$-body state $\RTE(V_n,\Delta t)|\psi_N\rangle^n$ 
approximately corresponds to the state $\RTE(\textsub V{KS}[\rho(t)],\Delta t)|\psi_N\rangle$ (see Appendix \ref{seca:trace} for derivation).
Therefore, $\RTE(\textsub V{KS}[\rho(t)],\Delta t)$ can be approximately implemented by applying $\exp(-\ii \Delta t V_n)$ on multiple copies as shown in \FIG{fg:circscf} (a). 
Here, the ``\trash'' denotes ``tracing out'' operation, which does not correspond to any physical action but simply means that the qubits are not used anymore.
In this way, we have automatically included the $\rho$-dependence of $\textsub V{KS}$ into ATE, but importantly, without ever requiring the expensive density readout and Hamiltonian re-encoding.
When the summation in \EQ{eq:mfn} is truncated at $n=2$, the full circuit for performing SCF within ATE is shown in \FIG{fg:circscf} (b).
Here, due to the no-cloning theorem, each copy is prepared individually and therefore we need a total of $2^{\textsub N{ATE}}$ copies to propagate $\textsub N{ATE}$ time steps.
At each step, only a half of the copies survive, and the others are traced out. 
Such exponential dependence of the number of qubits on $\textsub N{ATE}$ is evidently the major drawback of this algorithm and $\textsub N{ATE}$ is practically limited to rather small numbers.
However, because the density readout is avoided in this way, a reduction in total computation time can be expected.

\section{Numerical demonstrations}

In this section, we perform DFT calculations of several small systems using the methods proposed above to demonstrate their application potential.
With the first-quantized schemes, the bare Coulomb potential can lead to a number of problems.
Most notably, the singularity at the origin heavily makes its implementation very difficult \cite{huang2025approximate} and also affects the accuracy of ST decomposition used in ATE circuit and forces the maximum allowed time step $\Delta t$ to be very small, effectively increasing the computational cost.
With this consideration, we employ a GTH-type norm-conserving pseudopotential \cite{goedecker1996separable} in our calculations.
The pseudopotentials generally contain non-local terms, but their quantum implementation is still immature. 
Therefore, we only consider materials consisting of H and Li in the following examples, where relatively accurate results can be obtained with only local terms only.
All the calculation results here are obtained by simulating noise-free quantum circuits on classical computers, and the number of measurement shots are taken to be infinite (i.e., the results are directly calculated from the probability distribution).

\subsection{L\lowercase{i}H molecule: Harris functional and variational solution}\label{sec:lihHarris}

\begin{figure}
    \includegraphics[scale=0.55]{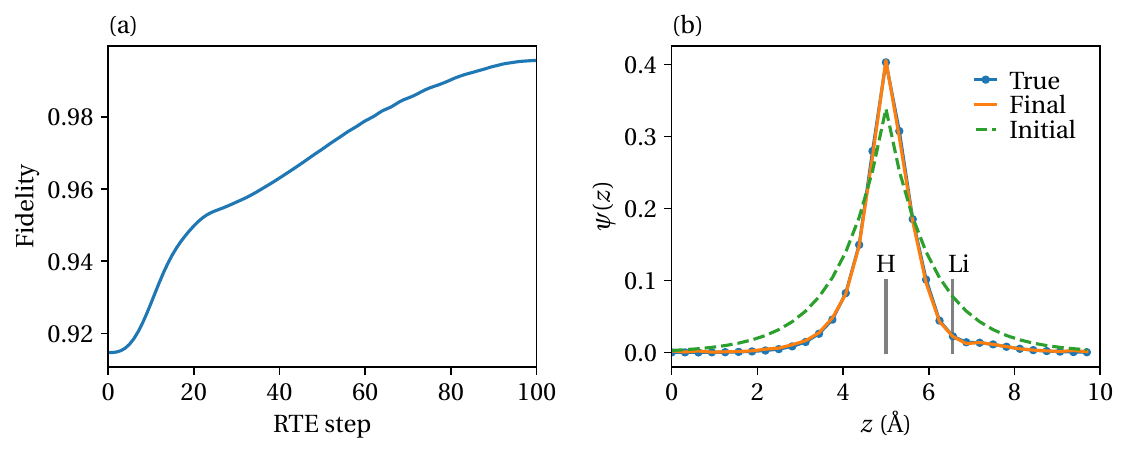}
    \caption{(a) The fidelity of wavefunction of LiH molecule with the true wavefunction during the ATE. (b) Wavefunctions along the line connecting H and Li atom.}\label{fg:lihpsi}
\end{figure}
\begin{figure}
    \includegraphics[scale=0.55]{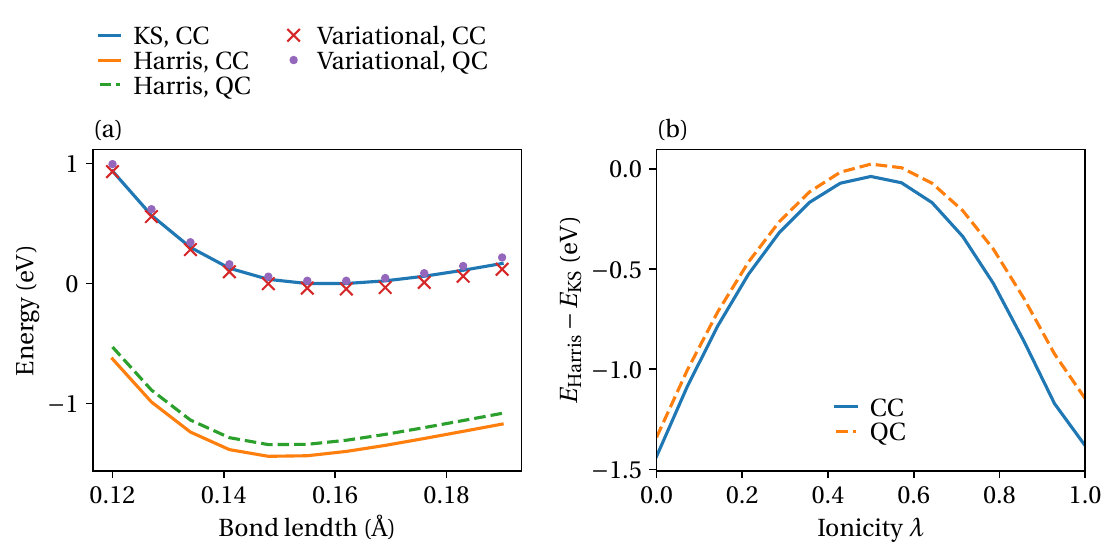}
    \caption{(a) Energy profile of LiH molecule with different bond lengths calculated by various methods. ``QC'' and ``CC'' denote quantum computation and classical computation (the true result), respectively. (b) Difference between Harris and KS total energy at Li--H distance \SI{1.55}{\angstrom} with different ionicity. }\label{fg:liherg}
\end{figure}

We first consider a LiH molecule for demonstration of the Harris functional with LDA exchange-correlation.
The molecule is placed in a $(\SI{10}{\angstrom})^3$ cubic box and each dimension is discretized using 32 equidistant points ($3\times5$ qubits in total), corresponding to a planewave cutoff of \SI{385}{eV}.
Since there is only one orbital to be calculated, the band qubits are not required.
For the initial wavefunction of the ATE, we have chosen a Slater-type orbital which is close to the real atomic orbitals and can be prepared with circuit depth $\mathcal O (\textsub n{grid})$ \cite{klco2020minimally}
\[\psi^0(\bm r)=\mathcal N \exp[-q(|x-x_0|+|y-y_0|+|z-z_0|)]\]
with $\mathcal N$ the normalization factor, $q=\SI{0.5}{\angstrom^{-1}}$ and $(x_0,y_0,z_0)$ the position of H atom.
The total time $\textsub tf$ and time step $\Delta t$ are important hyper-parameters for ATE which should be carefully tuned.
Adiabacity requires a large $\textsub tf$ while a relatively small $\Delta t$ is required for the accuracy of ST decomposition. Therefore, the higher required computational accuracy, the larger number of total ATE steps ($\textsub N{ATE}=\textsub tf/\Delta t$) is required. For a fixed $\textsub N{ATE}$, $\textsub tf$ (or $\Delta t$) must be chosen to balance these two types of error.
For this system, we have chosen $\textsub N{ATE}=100$ and $\textsub tf=\SI{20}{au}$, and have set $E_0=\SI{1}{au}$ in \EQ{eq:h0}.
At the equilibrium Li--H bond length $\SI{1.55}{\angstrom}$, this setting can bring the fidelity $|\langle\textsub\psi{true}|\textsub \psi{ATE}\rangle|^2$ from the initial 91.5\% to 99.6\% and final wavefunction is almost identical to the true wavefunction, as shown in \FIG{fg:lihpsi}.
The green dashed line in \FIG{fg:liherg} (a) shows the total energy calculated with Hadamard test following ATE at different Li--H bond lengths.
It is found that the quantum algorithm successfully reproduced the true energy surface within \SI{0.1}{eV}.

Comparing the energy surface obtained from Harris functional and full SCF KS result [orange and blue lines in \FIG{fg:liherg} (a)], 
we find that although the absolute energy from Harris functional is shifted, the relative energies at different bond lengths are accurately captured.
Additionally, $\textsub E{Harris}$ is lower than $\textsub E{KS}$ as almost always the case in practice.
As mentioned in subsection \ref{sec:harrisvar}, this properties of Harris function allows a variational approach on the input electronic density to improve the accuracy.
Here, we consider the electron transfer from Li to H, and model the real electronic density by
\[\textsub \rho{in}(\lambda)=(1+\lambda)\textsub \rho{isolated H}+(1-\lambda)\textsub \rho{isolated Li},\]
where the parameter $\lambda$ has a physical meaning of ionicity and $\lambda=0$ and $1$ corresponds to zero and full electron transfer, respectively.
It is indeed observed that the calculated total energy exhibits a maximum between $\lambda=0$ and $1$ as shown in \FIG{fg:liherg} (b) for bond length \SI{1.55}{\angstrom} as an example, and the energy surface obtained by this variational method is very close to the KS result as shown by the red cross in \FIG{fg:liherg} (a).
Such process can evidently also be performed on quantum computers.
We have found that as the $\lambda$ increases, as the final wavefunction shifts towards to the Li atom due to increased Coulomb repulsion by the increased charge on the H, it is better also to move the center of initial wavefunction.
Here, by taking the initial wavefunction center to be $\frac\lambda 2\textsub{\bm{R}}H+\left(1-\frac\lambda 2\right)\textsub{\bm{R}}{Li}$ where $\textsub{\bm{R}}H$ and $\textsub{\bm{R}}{Li}$ denotes the position of H and Li atom, respectively, our quantum algorithm also well reproduced the true energy surface as shown by the purple dots in \FIG{fg:liherg} (a).

\subsection{L\lowercase{i} metal under Harris functional: $\bm k$-point sampling and band structure}

As a further example on Harris functional, we consider the bulk BCC Li metal, focusing on the possibility of calculating metallic system with $\bm k$-point sampling and band structures on quantum computers.
We employ a conventional BCC cell consisting of 2 atoms with lattice constant $a=\SI{3.5}{\angstrom}$, discretized using 4 qubits on each dimension.

\begin{figure}
    \includegraphics[scale=0.55]{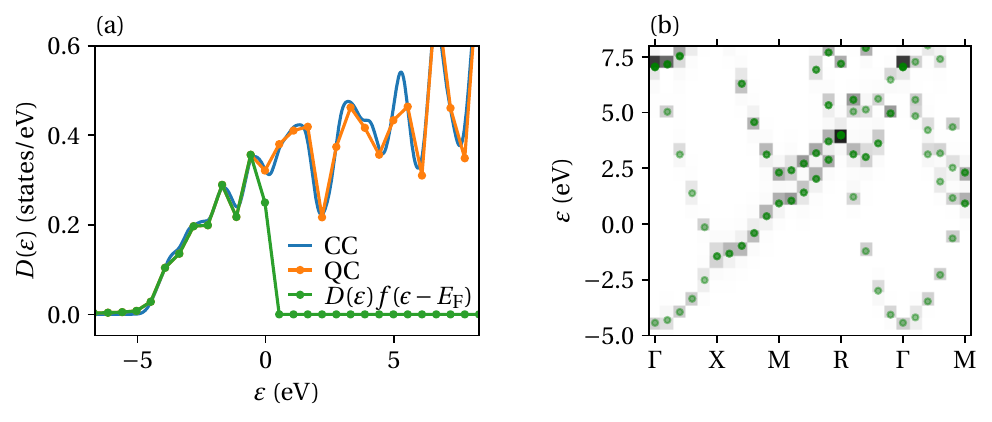}
    \caption{(a) Calculated DOS of BCC Li using the quantum (orange) and classical (blue) algorithms. The DOS of the quantum result multiplied by the smearing function $f$ is shown in green. (c) The band structure of Li metal. The green dots show the result from classical algorithm and the color map is the probability distribution in the measurement of the QPE.}\label{fg:limetal}
\end{figure}

While the system only contains 2 electrons, due to the fact the bulk Li is metallic, the $\textsub N{band}$ must be significantly increased from $\frac{\textsub N{elec}}2=1$.
Here, we initialize the wavefunctions using planewaves which can be implemented by QFT, $\psi_i^0(\bm r)=\frac1{\sqrt{V}}\exp\left(\frac{2\pi\ii}{a}\bm K_i\cdot \bm r\right)$, where $\bm K_i$ includes all integral vector with $K_i^2\le2$ (19 in total).
We have found that the $\textsub N{band}$ cannot be further reduced to obtain converged results.
In the current calculation with unoccupied states, QPE is required to compute the DOS and then the band energy.
Though for this system, 50 ATE steps with $\Delta t=\SI{0.15}{au}$ can already yield accurate DOS (see below), 
the achievable energy resolution is limited by the duration of Hamiltonian simulation in QPE as $\delta\varepsilon=\frac{2\pi}{\textsub N{QPE}\Delta t}$, and to reach $\delta\varepsilon\approx\SI{0.02}{au}$, $\textsub N{QPE}\approx 2^{11}$ (11 QPE working qubits) would be necessary,
which is actually the heaviest part of the whole calculation.

For the total-energy calculations, we have sampled the Brillouin zone using a $9\times9\times9$ $\bm k$-point mesh which contains 35 irreducible $\bm k$ points.
The orange line in \FIG{fg:limetal} (a) shows the DOS averaged on all the $\bm k$ points computed by the quantum algorithm, which well reproduced the results from conventional classical algorithm.
The Fermi level $\EF$ is then calculated using \EQ{eq:fermi} where the Gaussian smearing $f(x)=\frac12\erfc\left(\frac{x}{\sigma}\right)$ is employed with $\sigma=\SI{0.05}{eV}$, and $D(\varepsilon)f(\varepsilon-\EF)$ is shown by the green line in \FIG{fg:limetal} (a). 
The band energy is then calculated with \EQ{eq:fermie} and we have found that the band energy calculated using quantum algorithm agrees with the correct value within \SI{0.1}{eV/atom}.

Moreover, we have also calculated the DOS along the high-symmetry path $\Gamma(0,0,0)\to\text{X}\left(0,0,\frac12\right)\to\text{M}\left(0,\frac12,\frac12\right)\to\text{R}\left(\frac12,\frac12,\frac12\right)\to\Gamma\to \text{M}$ in the Brillouin zone.
We can then plot the band diagram by stacking these calculated DOS as shown by the color map in \FIG{fg:limetal} (b), which also well reproduced the results from classical results (green dots in the figure).
It should be noted that since we have used a 2-atom conventional BCC cell, the energy bands are folded and different from that calculated using the primitive cell which only contains 1 atom.
This demonstrates that our quantum algorithm is also useful to calculate the DFT band structures.

\section{L\lowercase{i}H molecule: SCF calculation using copies}

We finally exam the performance of SCF calculations of LiH molecule using copies of wavefunctions.
Here, the size of the simulation cell and number of grids on each dimension is halved due to the cost of simulation, and the summation in \EQ{eq:mfn} is truncated at $n=2$.
Additionally, as the non-smoothness of the bare Coulomb and delta function potential in \EQ{eq:v2} will affect the accuracy of ST decomposition, we have instead used a modified $v_1$ and $v_2$ in \EQs{eq:v1} and \ref{eq:v2} as 
\begin{align}
v_2(\bm r_1,\bm r_2)&{}=\frac1{|\bm r_1-\bm r_2|}\erf\left(\frac{|\bm r_1-\bm r_2|}{\sqrt 2 \textsub rc}\right)+\frac{2c_2}{\Delta V}\exp\left(-\frac{|\bm r_1-\bm r_2|^2}{2 \textsub rc^2}\right)\\
v_1(\bm r_1)&{}=\textsub V{ext}(\bm r_1)+\textsub V{H}[\textsub \rho{in}](\bm r_1)+\textsub V{XC}[\textsub \rho{in}](\bm r_1)-\sum_{\bm r_2} v_2(\bm r_1,\bm r_2)\textsub\rho{in}(\bm r_2)\Delta V,
\end{align}
where $\textsub rc=\SI{0.3}{au}$ and $\textsub \rho{in}$ is the input density for Harris functional.
In other words, we have smoothed the $v_2$ and added a correction term into $v_1$ so that the $\textsub V{KS}$ remains unchanged at $\textsub \rho{in}$.
We have found that this treatment results in final electron density virtually identical to that calculated using the exact KS scheme.

\begin{figure}
    \includegraphics[scale=0.55]{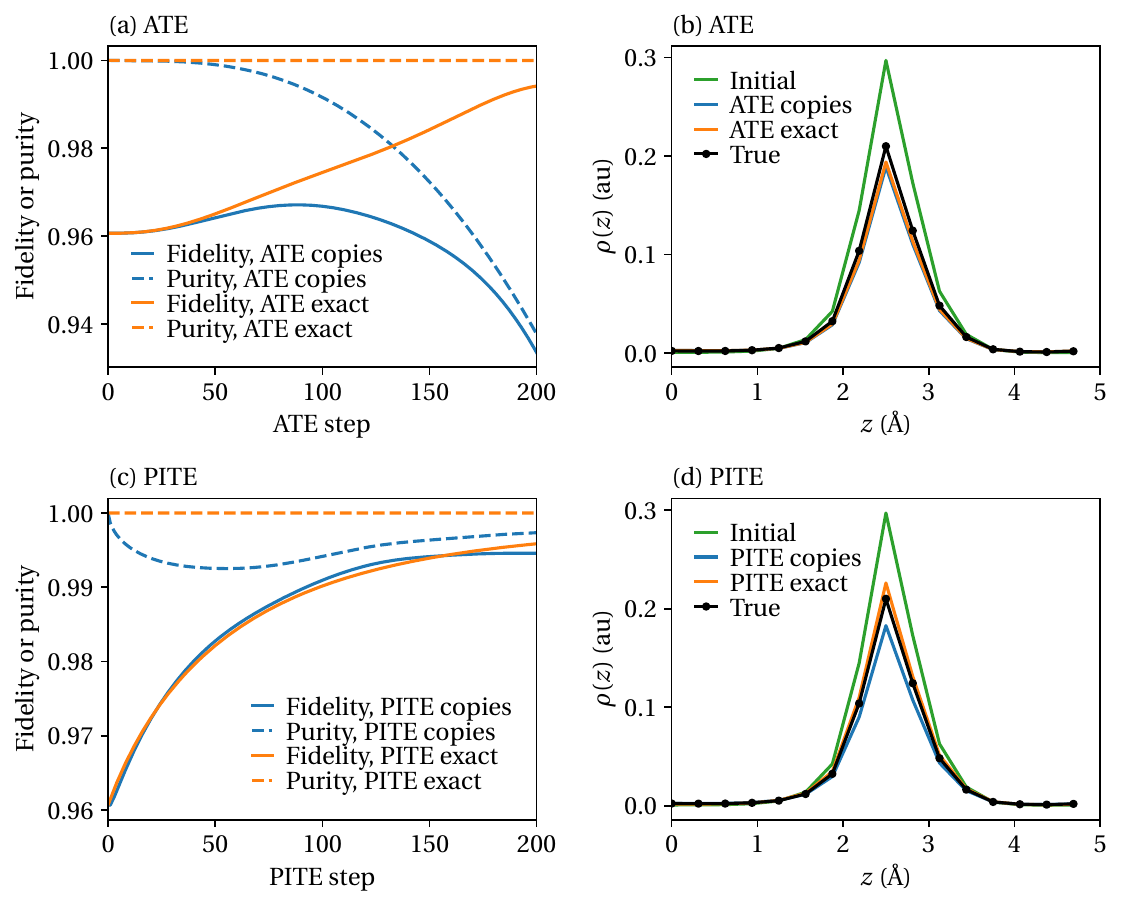}
    \caption{(a) Fidelity and state purity during ATE for SCF calculation of LiH molecule. The orange curves are the results of exact implementation of $\RTE(\textsub V{KS}[\rho(t)],\Delta t)$ by density readout, and the blue curves are the results for approximated implementation with multiple copies with wavefunctions. (b) Initial and true electronic density of LiH along the line connecting H and Li atom, and those calculated using ATE. (c)(d) The same as (a) and (b), respectively, but calculated using probabilistic imaginary time evolution (PITE) instead of ATE.}\label{fg:lihcopy}
\end{figure}

As the approximated $\RTE(\textsub V{KS}[\rho(t)],\Delta t)$ operator in \FIG{fg:circscf} (a) involves tracing out operation and is not strictly unitary, the quantum states will evolve into mixed state due to entanglement with other copies. 
This type of error can be monitored by the purity of the quantum state which we define as the largest eigen value of the one-copy reduced density matrix $\rho_{\bm r\bm r'}$.
Pure states will have purity 1 and any mixed states will have smaller purities. 
\FIG{fg:lihcopy} (a) shows the evolution of the state purity and fidelity $\langle\textsub \psi{true}|\rho_{\bm r\bm r'}|\textsub \psi{true}\rangle$ during the ATE with 200 steps and $\Delta t=\SI{0.1}{au}$ starting from the wavefunction calculated under Harris functional.
The blue and orange lines are the calculated results when the $\RTE(\textsub V{KS}[\rho(t)],\Delta t)$ operator is approximated using copies and implemented exactly (by reading out the density readout), respectively.
It is found that with the exact implementation of $\RTE(\textsub V{KS}[\rho(t)],\Delta t)$, high fidelity over 99\% can be reached in this case.
On the other hand, with approximated implementation using copies, the state purity gradually deteriorates during approximated ATE, preventing high fidelity to be reached (note that the fidelity cannot exceed the purity).
We have found that while increasing $\textsub N{ATE}$ with fixed $\textsub tf$ does improve the purity and fidelity, but the effect is somewhat limited. 
For example, to reach 98\% fidelity, $\textsub N{ATE}\sim 1000$ will be necessary.
However, the final electronic density is less sensitive to the approximation brought by the tracing out operation, and shows relatively smaller error compared to the fidelity as shown in \FIG{fg:lihcopy} (b).

For this system, since there is only one orbital to be computed, the PITE algorithm (its evolution for mixed states is given in Appendix \ref{seca:pite}) can also be applied.
Because the exact imaginary time evolution can decay all the excited states, it can evolve mixed state into pure states, which is not achievable in ATE.
When PITE is approximately implemented using copies, while purity still goes below unity, 
we have found that its deterioration can be automatically regulated, resulting in a much higher fidelity compared with ATE, as shown in \FIG{fg:lihcopy} (c).
This indicates that the SCF calculation with copies are better to be paired with PITE,
and more suitable in the case when only the lowest eigen states are required, such as OFDFT calculations.

\section{Discussion: Computational cost}\label{sec:discussion}

Achieving high-speed DFT calculation is crucial for application to large systems or high throughput screening.
In classical algorithms, the computational bottleneck lies on the diagonalization of the KS Hamiltonian and typically scales with $\mathcal O(\textsub N{atom}^3)$, making the affordable system sizes rather limited.
While by exploiting the nearsightedness \cite{prodan2005nearsightedness} of the electronic system, $\mathcal O(\textsub N{atom})$ scaling can be achieved \cite{galli1992large,yang1991direct}, such methods typically have very large pre-factors and are more difficulty to control the accuracy \cite{suryanarayana2017nearsightedness}. 
On quantum computers, the problem of diagonalization is not automatically solved because most of the quantum solvers can only compute the lowest eigen state efficiently.
With our method of QE and ATE, however, not only can all the occupied KS orbitals (instead of only the lowest one) be obtained at once without ever reading out the electronic density, but also the computational cost, specifically the required $\textsub N{ATE}$ and measurement shots to obtain the prescribed energy accuracy per atom, are both $\mathcal O(1)$ and do not increase with the system size.
These properties make it highly efficient compared to existing quantum solvers for excited states such as QLanczos or VQD \cite{motta2020determining,higgott2019variational}  for DFT calculation, where the expensive readout of the KS orbitals (not even density) would be required to obtain higher KS orbitals.
There is another category of the quantum algorithms that does not explicitly compute the KS orbitals but use eigenvalue filtering based on QSVT to project an input state into the subspace spanned by the occupied KS orbitals \cite{ko2023implementation},
and have the same computational cost scaling as our methods.
However, such algorithms uniformly sample the whole $\textsub N{grid}$-dimensional Hilbert space which is much larger than the occupied subspace ($\frac{\textsub N{grid}}{\textsub N{elec}}$ typically $10^3$ order), resulting a very low success rate in eigenvalue filtering.
Moreover, as the largest eigenvalue of $\textsub H{KS}$ is also much larger than the highest occupied orbital ($\frac{\max_{\text{all}}\varepsilon_i-\min\varepsilon_i}{\max_{\text{occupied}}\varepsilon_i-\min\varepsilon_i}$ is also typically sub-$10^3$ order), the implementation of eigenvalue filtering itself is very hard and requires $\sim 10^3$-order polynomial for QSVT.
In contrast, our method can generate the occupied subspace only with $10^2$ ATE steps, and therefore is expected to be dozens of times faster.

In order to run the SCF loop, the electronic density is directly readout to update the $\textsub H{KS}$ in all of the existing quantum algorithms \cite{ko2023implementation,senjean2023toward,nishiya2024orbital}, which would require $\mathcal O(\textsub N{grid})$ shots and is inefficient for large systems. 
However, by using Harris functional, while the accuracy is slightly degraded compared to the full SCF calculation, absolutely no density readout is needed, and this should be the method of choice for a majority of applications.
If improved accuracy is required, our variational method with Harris functional can be employed, which still requires no density readout and therefore can provide significant speed boost compared with the traditional methods.
Finally, for performing the full SCF KS-DFT calculations, while our method of using copies can be employed to avoid the density readout, it unfortunately should be regarded as a galactic algorithm as it requires exponential number if qubits.
Considering that solving the exact solution of the many-electronic system on quantum computer only costs $\mathcal O(\textsub N{atom})$ in circuit depth \cite{huang2025approximate},
instead of sticking to the DFT calculation, it might be more efficient to just solving the exact solution in practice if very accurate results are called for \cite{babbush2023quantum}. 
However, we believe the method with copies has its own theoretical value and may be adapted to solve other related problems efficiently.

\section{Conclusion}

In conclusion, we have proposed a qubit-efficient encoding scheme to represent the wavefunctions for DFT calculations, and all KS orbitals can be solved simultaneously by the ATE algorithm.
We found that Harris functional synergies well with quantum computers by using our algorithms, 
where the band energy can be exponentially accelerated by quantum computers and 
the heavy density readout is completely unnecessary.
While the Harris functional is slightly less accurate compared with the full KS DFT, its accuracy may be improved by a variational method that is also suitable for quantum computers.
We have also discussed a quantum algorithm for SCF calculation using multiple copies but can again avoid the heavy density readout during SCF.
From numerical tests, we find our methods to be highly efficient and may be applied in the near future to enable faster DFT calculations on quantum computers.

\begin{acknowledgments}
This work was partially supported by the Center of Innovations for Sustainable Quantum AI (JST Grant Number JPMJPF2221). The computation in this work has been done using the facilities of the Supercomputer Center, the Institute for Solid State Physics, the University of Tokyo (ISSPkyodo-SC-2026-Ea-0014).
\end{acknowledgments}

\bibliography{qdft.bib}
\appendix
\section{Level crossing}\label{seca:cross}
We here prove the assertion in subsection \ref{sec:cross}.
When level crossings occur during ATE, the \EQ{eq:qpestate} becomes
\[
|\text{QPE}\rangle=\sum_{kij\bm r}\sqrt{\frac{\Delta V}{\textsub N{band}}}\mathcal U_{ij}\psi_j^{\text{KS}}(\bm r) F_{jk} |k\rangle|i\rangle|\bm r\rangle,
\]
and \EQ{eq:qpeprob} will be
\[
\Pr(k)=\frac{1}{\textsub N{band}}\sum_{ij}^{\textsub N{band}} |\mathcal U_{ij}|^2|F_{jk}|^2.\] 
If \EQ{eq:uijsqr} is satisfied, by the unitarity of $\mathcal U$, we have for any $j$ occupied
\[\sum_{i}^{\textsub N{band}} |\mathcal U_{ij}|^2=1-\sum_{i>\textsub N{band}} |\mathcal U_{ij}|^2\approx 1,\]
thus the DOS of all the \emph{occupied orbitals} are not affected and the calculated band energy remains correct.

\section{The effect of tracing out}\label{seca:trace}
We here derive the reduced density matrix by tracing out in subsection \ref{sec:trace}.
Let $\rho_{i_\alpha i_\alpha'\bm r_\alpha \bm r_\alpha'}$ be the density matrix of the $\alpha$-th copy.
The action of $\exp(-\ii \Delta t V_n)$ on element of unentangled $n$-copy density matrix is
\[\rho_{i_1i_1'\bm r_1 \bm r_i'}\prod_{\alpha=2}^n\rho_{i_\alpha i_\alpha'\bm r_\alpha \bm r_\alpha'}
\to
\rho_{i_1i_1'\bm r_1 \bm r_i'}\prod_{\alpha=2}^n\rho_{i_\alpha i_\alpha'\bm r_\alpha \bm r_\alpha'}\exp[-\ii \Delta t \textsub N{elec}^{n-1} [v_n(\bm r_1,\bm r_2,\cdots\bm r_n)-v_n(\bm r_1',\bm r_2',\cdots\bm r_n')]],\]
After tracing out all other copies, the reduced density matrix of the first copy is 
\[\label{eq:trace}
\begin{split}
\rho_{i_1i_1'\bm r_1 \bm r_i'}^{\text{reduced}}&{}=\rho_{i_1i_1'\bm r_1 \bm r_i'}\sum_{\{i_\alpha\bm r_\alpha\}}\prod_{\alpha=2}^n\rho_{i_\alpha i_\alpha\bm r_\alpha \bm r_\alpha}\exp\{-\ii \Delta t \textsub N{elec}^{n-1} [v_n(\bm r_1,\bm r_2,\cdots\bm r_n)-v_n(\bm r_1',\bm r_2,\cdots\bm r_n)]\}\\
&{}=\rho_{i_1i_1'\bm r_1 \bm r_i'}\left\{1-\ii \Delta t (\Delta V)^{n-1} \sum_{\{\bm r_\alpha\}}\prod_{\alpha=2}^n\rho(\bm r_\alpha)[v_n(\bm r_1,\bm r_2,\cdots\bm r_n)-v_n(\bm r_1',\bm r_2,\cdots\bm r_n)]\right\}+\mathcal O(\Delta t^2)\\
&{}=\rho_{i_1i_1'\bm r_1 \bm r_i'}\exp\left\{-\ii \Delta t \left(\textsub V{KS}^{(n)}[\rho](\bm r_1)-\textsub V{KS}^{(n)}[\rho](\bm r_1')\right)\right\}+\mathcal O(\Delta t^2)\\
\end{split}\]
Here, we have used the fact that $\sum_{i_\alpha}\rho_{i_\alpha i_\alpha\bm r_\alpha \bm r_\alpha}=\frac{\Delta V}{\textsub N{elec}}\rho(\bm r_\alpha)$ where $\rho(\bm r_\alpha)$ denotes the electronic density at $\bm r_\alpha$.

\section{PITE of mixed states}\label{seca:pite}
The PITE circuit is given in Ref. \cite{kosugi2022imaginary}, and we here calculate its action to a mixed state represented by density matrix $\rho$.
The state of ancilla before controlled RTE is 
\[\begin{bmatrix}1\\0\end{bmatrix}
\xrightarrow{H}
\frac1{\sqrt 2}\begin{bmatrix}1\\1\end{bmatrix}
\xrightarrow{W\equiv \frac1{\sqrt2} \begin{bmatrix}1&-\ii \\ 1& \ii\end{bmatrix}}
\frac1{2}\begin{bmatrix}1-\ii\\1+\ii\end{bmatrix}.\]
The total density matrix is then \[\rho_A\otimes \rho=\frac 12 \begin{bmatrix}\rho & -\ii\rho \\ \ii\rho & \rho\end{bmatrix},\]
where $\rho_A$ and $\rho$ denote density matrix of ancilla and wavefunction qubits, respectively.
The 0-RTE ($\RTE$ denotes RTE operator) brings the density matrix to
\[\begin{bmatrix}\RTE \\ & \mathbb I\end{bmatrix}
\frac 12 \begin{bmatrix}\rho & -\ii\rho \\ \ii\rho & \rho\end{bmatrix}
\begin{bmatrix}\RTE^\dagger \\ & \mathbb I\end{bmatrix}
=\frac 12 \begin{bmatrix}\RTE\rho \RTE^\dagger & -\ii \RTE\rho  \\ \ii\rho \RTE^\dagger & \rho\end{bmatrix}.\]
then after 1-RTE$^\dagger$,
\[\begin{bmatrix}\mathbb I \\ &\RTE^\dagger \end{bmatrix}
\frac 12 \begin{bmatrix}\RTE\rho \RTE^\dagger & -\ii \RTE\rho  \\ \ii\rho \RTE^\dagger & \rho\end{bmatrix}
\begin{bmatrix}\mathbb I \\ & R\end{bmatrix}
=\frac 12 \begin{bmatrix}\RTE\rho \RTE^\dagger & -\ii \RTE\rho \RTE  \\ \ii \RTE^\dagger\rho \RTE^\dagger & \RTE^\dagger\rho \RTE\end{bmatrix}.\]
Finally, we apply $W^\dagger R_z=
\frac1{\sqrt2} \begin{bmatrix}1&1 \\ \ii& -\ii\end{bmatrix}
\begin{bmatrix}\ee^{-\ii\theta} \\ & \ee^{\ii\theta}\end{bmatrix}
=\frac1{\sqrt2}\begin{bmatrix}\ee^{-\ii\theta} & \ee^{\ii\theta} \\ \ii\ee^{-\ii\theta} & -\ii\ee^{\ii\theta}\end{bmatrix}
$ to ancilla, and obtain
\[\begin{split}&
\frac1{\sqrt2}\begin{bmatrix}\ee^{-\ii\theta} & \ee^{\ii\theta} \\ \ii\ee^{-\ii\theta} & -\ii\ee^{\ii\theta}\end{bmatrix}
\frac 12 \begin{bmatrix}\RTE\rho \RTE^\dagger & -\ii \RTE\rho \RTE  \\ \ii \RTE^\dagger\rho \RTE^\dagger & \RTE^\dagger\rho \RTE\end{bmatrix}
\frac1{\sqrt2}\begin{bmatrix}\ee^{\ii\theta} & -\ii\ee^{\ii\theta} \\ \ee^{-\ii\theta} & \ii\ee^{-\ii\theta}\end{bmatrix}\\
&{}= \frac14\begin{bmatrix} \RTE\rho \RTE^\dagger  -\ii\ee^{-2\ii\theta} \RTE\rho \RTE + \ii \ee^{2\ii\theta} \RTE^\dagger\rho \RTE^\dagger + \RTE^\dagger\rho \RTE & -\ii \RTE\rho \RTE^\dagger  +\ee^{-2\ii\theta} \RTE\rho \RTE +\ee^{2\ii\theta} \RTE^\dagger\rho \RTE^\dagger + \ii \RTE^\dagger\rho \RTE \\
    \ii \RTE\rho \RTE^\dagger  +\ee^{-2\ii\theta} \RTE\rho \RTE +  \ee^{2\ii\theta} \RTE^\dagger\rho \RTE^\dagger -\ii \RTE^\dagger\rho \RTE &  \RTE\rho \RTE^\dagger  +\ii \ee^{-2\ii\theta} \RTE\rho \RTE -\ii\ee^{2\ii\theta} \RTE^\dagger\rho \RTE^\dagger + \RTE^\dagger\rho \RTE
        \end{bmatrix}.
\end{split}\]
Thus the un-normalized density matrix after measurement of ancilla with result 0 is \[\RTE\rho \RTE^\dagger  -\ii\ee^{-2\ii\theta} \RTE\rho \RTE + \ii \ee^{2\ii\theta} \RTE^\dagger\rho \RTE^\dagger + \RTE^\dagger\rho \RTE.\]

\end{document}